\documentclass[10pt,conference]{IEEEtran}
\IEEEoverridecommandlockouts
\usepackage{cite}
\usepackage{amsmath,amssymb,amsfonts}
\usepackage{algorithmic}
\usepackage{graphicx} 
\usepackage{booktabs}
\usepackage{textcomp}
\usepackage{subcaption}
\usepackage{tcolorbox}
\usepackage{amsmath}
\usepackage{amssymb}
\usepackage{enumerate}
\usepackage{enumitem}
\usepackage{alltt}
\usepackage{url}

\usepackage{listings}
\lstdefinelanguage{diff}{
  language=Java,
  morecomment=[f][\color{blue}]{@@},     
  morecomment=[f][\color{red!60!black}]-,         
  morecomment=[f][\color{green!60!black}]+,       
  morecomment=[f][\color{magenta}]{---}, 
  morecomment=[f][\color{magenta}]{+++},
}
\usepackage{stackengine}
\usepackage{boxhandler}

\newcommand{\OurTool}{\text{$\mu$\textsc{Bert}}}
\begin{document}
 \title{\OurTool: Mutation Testing using Pre-Trained Language Models}

\author{
\IEEEauthorblockN{Renzo Degiovanni}
\IEEEauthorblockA{SnT, University of Luxembourg, Luxembourg}
\and
\IEEEauthorblockN{Mike Papadakis}
\IEEEauthorblockA{SnT, University of Luxembourg, Luxembourg}
}

\maketitle         

\begin{abstract}
We introduce {\OurTool}, a mutation testing tool that uses a pre-trained language model (CodeBERT) to generate mutants. This is done by masking a token from the expression given as input and using CodeBERT to predict it. Thus, the mutants are generated by replacing the masked tokens with the predicted ones. We evaluate {\OurTool} on 40 real faults from Defects4J and show that it can detect 27 out of the 40 faults, while the baseline (PiTest) detects 26 of them. We also show that {\OurTool} can be 2 times more cost-effective than PiTest, when the same number of mutants are analysed. 
Additionally, we evaluate the impact of {\OurTool}'s mutants when used by program assertion inference techniques, and show that they can help in producing better specifications. Finally, we discuss about the quality and naturalness of some interesting mutants produced by {\OurTool} during our experimental evaluation.

\end{abstract}

\section{Introduction}

Mutation testing seeds faults using a predefined set of simple syntactic transformations, aka mutation operators, that are (typically) defined based on the grammar of the targeted programming language \cite{PapadakisK00TH19}. As a result, mutation operators often alter the program semantics in ways that often lead to unnatural code (unnatural in the sense that the mutated code is unlikely to  be produced by a competent programmer).

Such unnatural faults may not be convincing for developers as they might perceive them as unrealistic/uninteresting \cite{BellerWBSM0021}, thereby hindering the usability of the method. Additionally, the use of unnatural mutants may have actual impact on the guidance and assessment capabilities of mutation testing \cite{GopinathJG14}. This is because unnatural mutants often lead to exceptions, or segmentation faults, infinite loops and other trivial cases. 

To deal with this issue, we propose forming mutants that are in some sense natural; meaning that the mutated code/statement follows the implicit rules, coding conventions and generally representativeness of the code produced by competent programmers. We define/capture this naturalness of mutants using language models trained on big code that learn (quantify) the occurrence of code tokens given their surrounding code.

In particular, recent research 
has developed pre-trained models, such as 
CodeBERT~\cite{DBLP:conf/emnlp/FengGTDFGS0LJZ20}, 
using a corpus of more than 6.4 million programs, which could be used to generate natural mutants. 
Such pre-trained models have been trained to predict (complete) missing tokens (masked tokens) from token sequences. 
For example, given the masked sequence \texttt{int a = <mask>;}, CodeBERT predicts that \texttt{0}, \texttt{1}, \texttt{b}, \texttt{2},  and \texttt{10} are the (five) most likely tokens/mutants to replace the masked one (ordered in descending order according to their score -- likelihood). 

In view of this, we present {\OurTool}, a mutation testing tool that uses a pre-trained language model (CodeBERT) to generate mutants by masking and replacing tokens. 
{\OurTool} combines mutation testing and natural language processing  to form natural mutants. 
In contrast to resent research \cite{ChekamPBTS20, ma2021mudelta} that aims at mutant selection, {\OurTool} directly generates mutants without relying on any syntactic-based mutation operators. This approach is further appealing since it simplifies the creation of mutants and limits their number. 

\begin{figure*}[htp!]
\centering
\includegraphics[width=\textwidth]{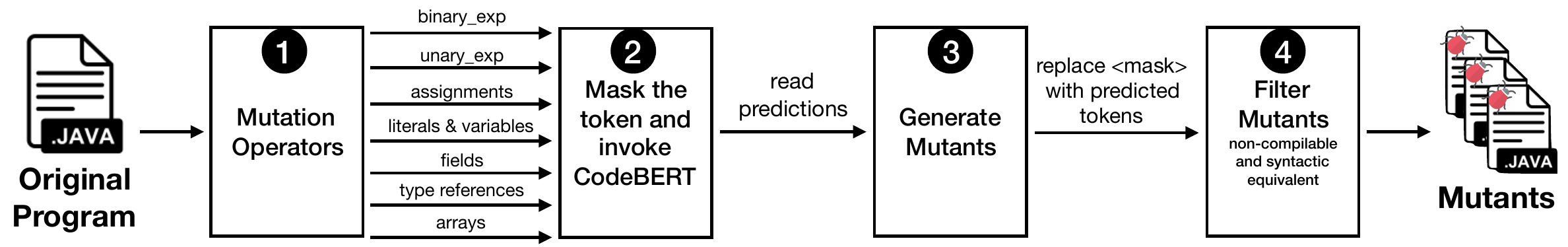}
\vspace{-1.4em}
\caption{{\OurTool} Workflow: (1) it parses the Java code given as input, and extracts the expressions to mutate according to the mutation operators; (2) it masks the token of interest and invokes CodeBERT; (3) it generates the mutants by replacing the masked token with CodeBERT predictions; and (4) it discards non-compilable and syntactic the same mutants.}
\label{fig:muBERT}
\end{figure*}

Although, there are many ways to tune {\OurTool} by considering mutants' locations and their impact, in our preliminary analysis, we seed faults in a brute-force way, similarly to mutation testing, by iterating every program statement and masking every involved token. In particular, we make the following steps: (1) select and mask one token at a time, depending on the type of expression being analysed; (2) feed CodeBERT with the masked sequence and obtain the predictions; (3) create mutants by replacing the masked token with the predicted ones; and (4) discard non-compilable and duplicate mutants (mutants syntactically the equal to original code). Figure~\ref{fig:muBERT} shows an overview of {\OurTool} workflow.

To show the potential of  {\OurTool} we perform a preliminary evaluation on the following two use cases:  

    \textbf{Fault Detection:} We focus on a mutation testing scenario and analyse the fault detection capabilities of suites designed to kill {\OurTool}'s mutants, and compare them with those of a popular mutation testing tool, i.e., PiTest~\cite{ColesLHPV16}. We consider a total of 40 bugs from Defects4J~\cite{defects4j} for 3 projects, namely Cli, Collections and Csv. Our results show that test suites guided by {\OurTool} finds 27 out of the 40 bugs, while PiTest's mutants helps in finding 26 out of the 40 bugs. 3 of the bugs found by {\OurTool} are not found  by PiTest, while 2 of the bugs found by PiTest are not found by {\OurTool}. Moreover, we show that  {\OurTool} is (up to 100\%) more cost-effective than PiTest.
    
    \textbf{Assertion inference:} We study the usefulness of {\OurTool}'s mutants in the context of program assertion inference techniques, that use mutants to rank and discard candidate assertions~\cite{DBLP:conf/issta/JahangirovaCHT18} (typically, assertions that kill more mutants are preferred among others, and assertions not killing any mutant are discarded). In particular, we focus on the 4 cases recently reported in~\cite{MolinaDA2022} in which traditional mutation testing 
did not perform well. We show that {\OurTool} can complement and contribute with interesting mutants than can help in improving the quality of the assertions inferred.

Finally, we show examples of the mutants generated by {\OurTool} with interesting properties, demonstrating their differences from traditional mutation.


\section{Pre-Trained Language Models}

CodeBERT~\cite{DBLP:conf/emnlp/FengGTDFGS0LJZ20} is a powerful bimodal pre-trained language model that produces general-purpose representations for natural language, in six programming languages, including Java. 
It supports several tasks, such as, natural language code search and code documentation. Particularly, CodeBERT supports the Masked Language Modelling (MLM) task that consists of randomly masking some of the tokens from the input, and the objective is to predict the original tokens of the masked word based only on its context. To do so, CodeBERT uses multi-layer bidirectional Transformer~\cite{VaswaniSPULALP2017} to capture the semantic connection between the tokens and the surrounding code, meaning that the predictions are context-dependent (e.g. the same variable name, masked in different program locations, will likely get different predictions). 

Precisely, CodeBERT can be fed with sequences of up to 512 tokens (maximum sequence length supported) that include exactly one (1) masked token (\texttt{<mask>}). Hence, when fed with a masked sequence, CodeBERT will predict the 5 most likely tokens to replace the masked one. 
Despite the good precision of CodeBERT in reproducing the original (masked) token, {\OurTool} uses all the predicted tokens to introduce mutations in the original program. We argue that mutations introduced by {\OurTool} will be in some sense natural, since CodeBERT was pre-trained on a large corpus (near 6.4 million programs) and thus, the mutated statements will follow frequent/repetitive coding conventions and patterns produced by programmers learned by the pre-trained language model. 

It is worth noticing that {\OurTool} uses CodeBERT as a black-box, so it will benefit from any improvement that the pre-trained model can bring in the future, as well as, other language models (supporting MLM task) can be integrated. Perhaps more importantly, generative pre-trained language models simplify the creation and selection of mutants to a standard usage of the model.

\section{\OurTool: CodeBERT-based Mutant Generation}
{\OurTool} is an automated approach that uses a pre-trained language model (namely, CodeBERT) to generate mutants for Java programs. 
Figure~\ref{fig:muBERT} describes the workflow of {\OurTool} that can be summarised as follows:
\begin{enumerate}
    \item {\OurTool} starts by parsing the Java class given as input, and extracts the candidate expressions to mutate. 
    \item The mutation operators analyse and mask the token of interest for each java expression (e.g., the binary expression mutation will mask the binary operator), and then invoke CodeBERT to predict the masked token. {\OurTool} will try to feed CodeBERT with sequences covering as much surrounding context as possible of the expression under analysis (512 tokens maximum).
    \item {\OurTool} takes CodeBERT predictions, and generate mutants by replacing the masked token with the predicted tokens (5 mutants are created per masked expression).
    \item Finally, mutants that do not compile, or are syntactic the same as the original program (cases in which CodeBERT predicts the original masked token), are discarded.
\end{enumerate}

Our prototype implementation supports a wide variety of Java expressions, being able to mutate unary/binary expressions, assignment statements, literals, variable names, method calls, object field accesses, among others. 
This indicates that for the same program location, several mutants can be generated. For instance, for a binary expression like \texttt{a + b}, {\OurTool} will create (potentially 15) mutants from the following 3 masked sequences: \texttt{<mask> + b}, \texttt{a <mask> b}, and \texttt{a + <mask>}. Bellow we provide some examples that demonstrate the different mutation operators supported by {\OurTool}.

\subsection{Binary Expression Mutation}
Given $e = \texttt{<exp> <op> <exp>}$, a binary expression of a method $M$ in program $P$ to mutate, where $\texttt{<exp>}$ and $\texttt{<op>}$ denote a Java expression and a binary operator, respectively, {\OurTool} creates a new expression $e' = \texttt{<exp> \textbf{<mask>} <exp>}$ by replacing (masking) the binary operator $\texttt{<op>}$ with the special token \texttt{<mask>}. 
Then, a new method $M' = M[e \leftarrow e']$ is created that looks exactly as $M$, but  expression $e$ is replaced by masked expression $e'$. 
{\OurTool} invokes CodeBERT with the largest code sequence from method $M'$ that, includes $e'$ and, does not exceed the maximum sequence length (512 tokens). 
CodeBERT returns a set with the 5 predicted tokens ($t_1,\ldots,t_5$). 
Hence, {\OurTool} generates 5 mutants, namely $P_1,\ldots,P_5$, such that each mutant $P_i$ replaces the mutated operator $\texttt{<op>}$ by the predicted one $t_i$. That is, $P_i = P[e \leftarrow e_i]$, where $e_i = \texttt{<exp>}\ t_i\ \texttt{<exp>}$ and $i \in [1..5]$.  
Finally, $\OurTool$ discards non-compilable mutants, and those that are syntactic the same as the original program (i.e., when $\texttt{<op>} = t_i$).

\begin{figure}[htp!]
\vspace{-0.85em}
\centering
\includegraphics[width=.48\textwidth]{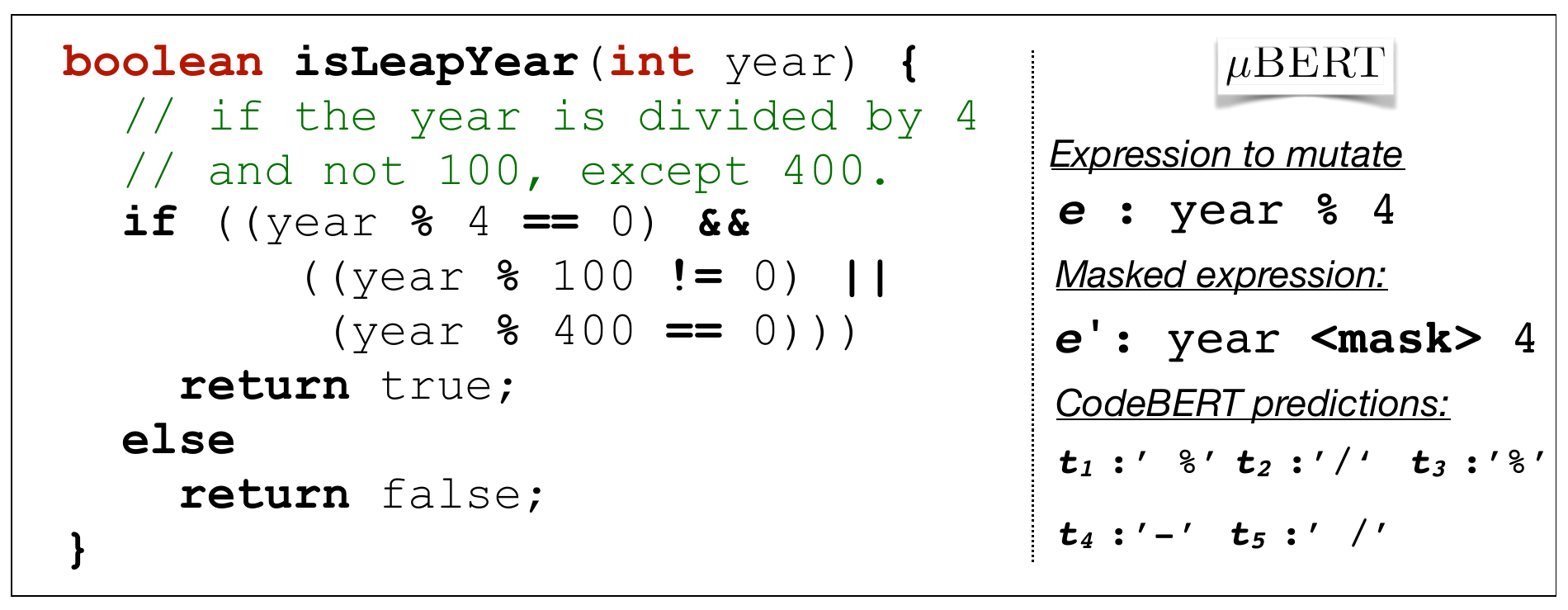}
\caption{{\OurTool}'s mutation operator for binary expressions.}
\label{fig:binary_example}
\vspace{-1.05em}
\end{figure}

Figure~\ref{fig:binary_example} shows one example of mutants that {\OurTool} can generated for binary expressions. Function \texttt{isLeapYear} returns true if a calendar year given as input is leap. 
One of the binary expressions to mutate is $e: \texttt{year \% 4}$. 
To do so, {\OurTool} masks binary operator \texttt{\%}, leading to masked expression $e': \texttt{year <mask> 4}$. 
The entire masked method is used to feed CodeBERT, for which it predicts the following 5 tokens: $t_1:\texttt{'\ \%'}$, $t_2:\texttt{'/'}$, $t_3:\texttt{'\%'}$, $t_4:\texttt{'-'}$ and $t_5:\texttt{'\ /'}$. 
First notice that tokens $t_1$ and $t_3$ only differs in a space and coincides with the original token, so these mutants will be discarded. 
Second, tokens $t_2$ and $t_5$ are the same, except the extra space in $t_5$, so only one will be used for generating the mutant. 
Finally, {\OurTool} produces 2 compilable mutants, based of the expressions $e_2: \texttt{year / 4}$ and $e_4: \texttt{year - 4}$.

\subsection{Unary Expression Mutation}
When dealing with unary expressions, {\OurTool} distinguishes two cases, depending if the operator appears before or after the expression (e.g.\texttt{++x} and \texttt{x--}). 
For the sake of simplicity, consider that $e= \texttt{<op><exp>}$ is the unary expression to mutate. 
Then, {\OurTool} will mask the operator token $\texttt{<op>}$, leading to masked expression $e'= \texttt{\textbf{<mask>} <exp>}$, and the masked sequence is then fed to CodeBERT.   
{\OurTool} takes CodeBERT predictions ($t_1,\ldots,t_5$) and creates mutants $P_1,\ldots,P_5$ by replacing the unary operator $\texttt{<op>}$ by the predicted tokens $t_i$. That is, $P_i = P[e \leftarrow e_i]$, where $e_i =  t_i\ \texttt{<exp>}$ and $i \in [1..5]$. 
Duplicated, syntactic the same and non-compilable mutants are finally discarded. 

\begin{figure}[htp!]
\centering
\includegraphics[width=.48\textwidth]{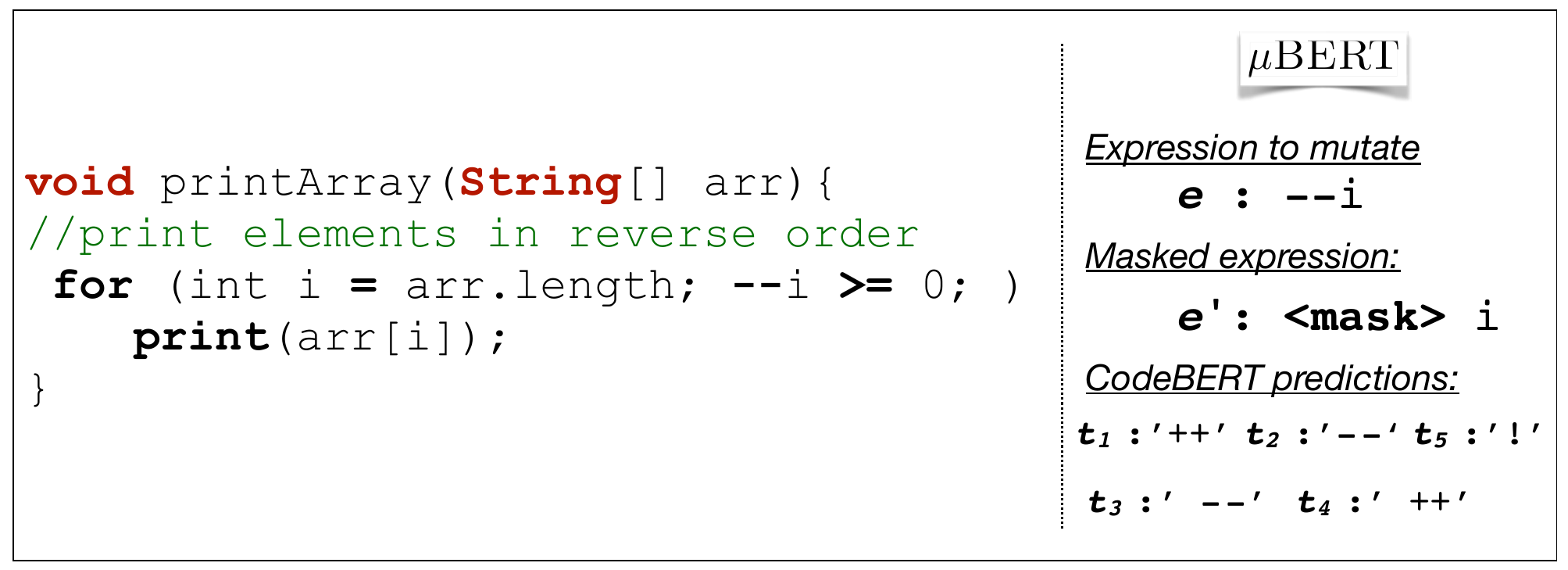}
\caption{{\OurTool}'s mutation operator for unary expressions.}
\label{fig:unary_example}
\vspace{-1.0em}
\end{figure}

Figure~\ref{fig:unary_example} shows an example of mutants that {\OurTool} can generate for unary expressions. 
Function \texttt{printArray} prints the elements of the array \texttt{arr} given as input in reverse order. Consider that {\OurTool} is going to mutate unary expression $e : \texttt{--i}$, for which it generates masked expression $e': \texttt{\textbf{<mask>} i}$ that is fed into CodeBERT. 
{\OurTool} receives the following predictions:
 $t_1:\texttt{'++'}$, $t_2:\texttt{'--'}$, $t_3:\texttt{' --'}$, $t_4:\texttt{' ++'}$ and $t_5:\texttt{'!'}$. 
{\OurTool} discards mutants syntactic the same as the original (tokens $t_2$ and $t_3$), and considers two candidate mutants ($t_1$ and $t_5$), but only mutation $t_1$ compiles (obtaining $e_1: \texttt{++i}$). 

\subsection{Literal and Variable Name Mutation}
This mutation is straightforward. 
For the sake of simplicity, consider that expression $e = \texttt{<cons>}$ to mutate is a literal (constant). 
{\OurTool} starts by masking $e$, leading to $e' = \texttt{\textbf{<mask>}}$ that is used to feed CodeBERT. 
{\OurTool} creates mutants $P_1,\ldots,P_5$ by replacing the mutated literal name by the predicted tokens (i.e., $P_i = P[e \leftarrow t_i]$ for $i \in [1..5]$). 

Consider again function \texttt{isLeapYear} from Figure~\ref{fig:binary_example}, where literal expression $e: \texttt{4}$ is the expression to mutate (from \texttt{year \% 4}). 
After replacing $e$ with mask token, CodeBERT returns the following 5 predictions: $t_1:\texttt{'4'}$, $t_2:\texttt{'100'}$, $t_3:\texttt{'400'}$, $t_4:\texttt{'10'}$ and $t_5:\texttt{'2'}$. 
Notice that, tokens $t_2$ and $t_3$ are present in the context of the mutated expression. 
Also note that first prediction ($t_1$) coincides with the original token, so it is discarded. Finally, {\OurTool} returns 4 compilable mutants, generated by replacing the masked token with predicted tokens $t_2$, $t_3$, $t_4$ and $t_5$.

\subsection{More Mutation Operators}
{\OurTool} is also able to mutate assignments, method calls, object field accesses, array reading and writing, and reference type expressions. 
Bellow we provide examples of the resulting masked sequences that {\OurTool} generates to mutate these kind of expressions. 
Following the same process already described before, {\OurTool} will generate the mutants by replacing the masked token with CodeBERT predictions. 
Notice that the shown predictions were observed during our experimentation, but these will likely change if are evaluated under different surrounding context. 

\begin{itemize}[wide=5pt,noitemsep,topsep=0pt]
\item For an assignment expression like $\texttt{avg += it\_result}$, {\OurTool} produces the masked expression $\texttt{avg \textbf{<mask>}= it\_result}$. Typical CodeBERT predictions are \texttt{+}, \texttt{-}, \texttt{*} and \texttt{/} leading to potential compilable mutants, e.g., $\texttt{avg -= it\_result}$.

\item In a method call expression, such as \texttt{children.add(c)} in Figure~\ref{fig:method_call_example}, {\OurTool} masks the method name, producing \texttt{children.\textbf{<mask>}(c)}. 
CodeBERT predicts the following method names: \texttt{add}, \texttt{addAll}, \texttt{push}, \texttt{remove} and \texttt{added}. {\OurTool} discards equally the same and non-compilable mutants, obtaining two mutants:  \texttt{children.push(c)} and \texttt{children.remove(c)}.

\begin{figure}[htp!]
\centering
\includegraphics[width=.48\textwidth]{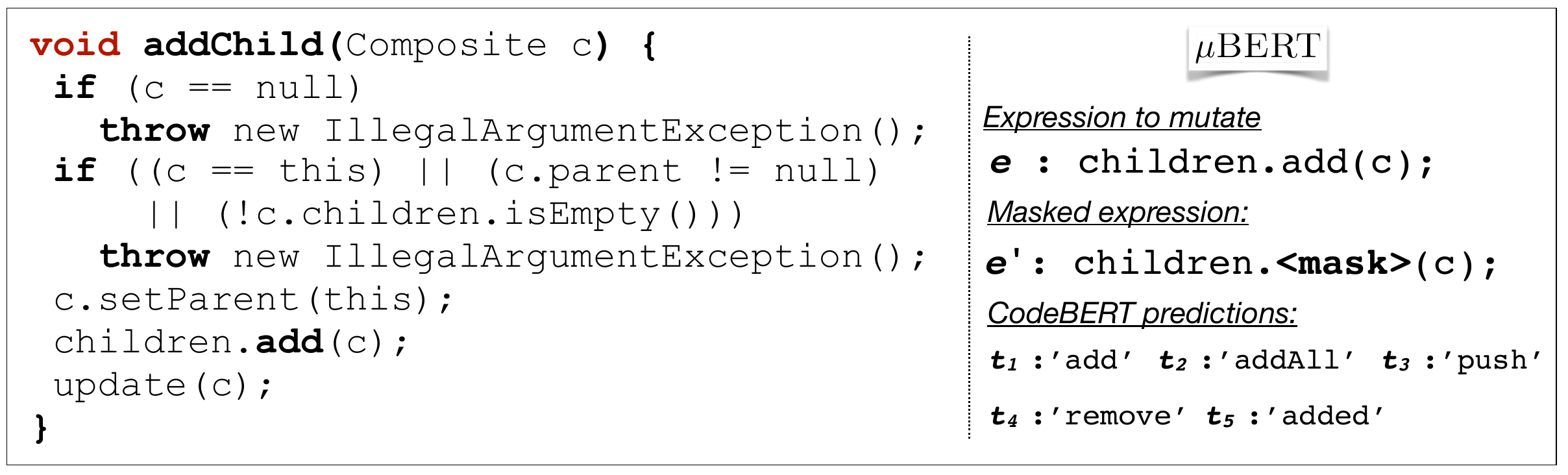}
\caption{{\OurTool}'s mutation operator for method calls.}
\label{fig:method_call_example}
\vspace{-1.0em}
\end{figure}

\item In expressions that access to particular object fields, {\OurTool} masks the object field name. For instance, for an expression like $\texttt{list.head = new\_node}$,  {\OurTool} produces the masked expression  $\texttt{list.\textbf{<mask>} = new\_node}$. CodeBERT predictions that we usually get cover \texttt{head}, \texttt{next}, \texttt{tail}, \texttt{last} and \texttt{first}.

\item In array reading (and/or writing) expressions, {\OurTool} masks the entire index used to access to the array. For instance, for the expression \texttt{arr[mid-1]} in Figure~\ref{fig:array_example}, {\OurTool} produces \texttt{arr[\textbf{<mask>}]}
masked expression. Then, CodeBERT predictions are \texttt{0}, \texttt{n}, \texttt{mid}, \texttt{1} and \texttt{low}, allowing to {\OurTool} generate 5 compilable mutants (variables \texttt{n}, \texttt{low} and \texttt{mid} are present in the context). 
It is worth noticing that the array name (\texttt{arr}) and the index expression \texttt{mid - 1} will be mutated by the variable name mutation operator and binary expression mutation operator, respectively.

\begin{figure}[htp!]
\centering
\includegraphics[width=.48\textwidth]{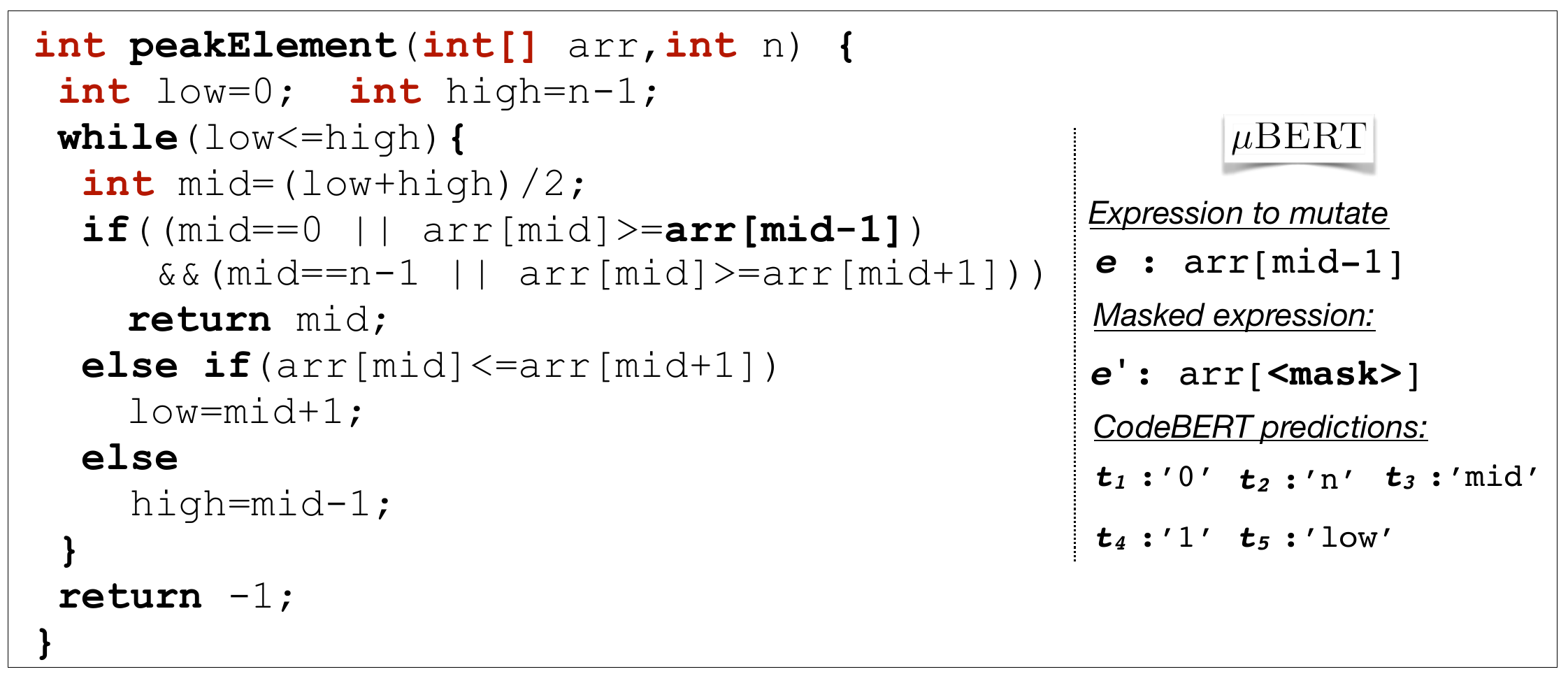}
\caption{{\OurTool}'s mutation operator for array expressions.}
\label{fig:array_example}
\vspace{-1.0em}
\end{figure}

\item In expressions that refers to some type, such as  {\small$\texttt{int number = (int)(Math.random() * 10)}$}, {\OurTool} masks that class name of the referred type. In this case, {\OurTool} produces the masked expression 
{\small$\texttt{int number = (int)(\textbf{<mask>}.random() * 10)}$}. 
For this example, predictions we obtained refer to \texttt{Math}, \texttt{random}, \texttt{Random} and \texttt{System}, leading to mutants such as 
{\small$\texttt{int number = (int)(Random.random() * 10)}$}.
\end{itemize}

\section{Research Questions}
We start our analysis by investigating the fault detection capabilities of test suites designed to kill {\OurTool}'s mutants. Thus, we ask:

\begin{description}
\item[RQ1] \emph{How effective are the mutants generated by {\OurTool} in detecting real faults? How does {\OurTool} compare with PiTest in terms of fault detection?}
\end{description}

To answer this question we evaluate the fault detection ability of test suites selected to kill the mutants produced by {\OurTool} and PiTest~\cite{ColesLHPV16}, our baseline. The fault detection ability is approximated by using a set of real faults taken from Defects4J~\cite{defects4j}.

Another application case of mutation testing regards the program assertion generation. In particular, using mutation testing for selecting and discarding assertions by program assertion inference techniques. In view of this, we ask:

\begin{description}
\item[RQ2] \emph{Is {\OurTool} successful in selecting ``good" assertions? How does it compare with PiTest?}
\end{description}

To answer this question we use a dataset composed by manually written  assertions (ground-truth) that was recently used for evaluating SpecFuzzer tool~\cite{MolinaDA2022}, a state-of-the-art specification inference technique. Particularly, we select 4 manually written assertions that were mistakenly discarded by SpecFuzzer, since they do not kill any mutant. We thus, investigate whether {\OurTool} can help in selecting these assertions and compare it with PiTest.

Finally, we qualitatively analyse some of the mutants generated with {\OurTool} and ask:

\begin{description}
\item[RQ3] \emph{Does {\OurTool} generates different mutants than traditional mutation testing operators?}
\end{description}

We showcase the mutants generated by {\OurTool} that help in detecting faults not found by PiTest, and mutants that help SpecFuzzer in preserving assertions from the ground-truth, that are discarded by mutants from PiTest.

\section{Experimental Setup}
\subsection{Faults and Assertions (Ground-truth)}

For the fault detection analysis, we use Defects4J~\cite{defects4j} v2.0,0, which contains the build infrastructure to reproduce (over 800) real faults for Java programs.  Every bug in the dataset consists of the faulty and fixed versions of the code and a developer’s test suite accompanying the project that includes at least one fault triggering test that fails in the faulty version and passes in the fixed one. 
Since this is a preliminary evaluation, we target projects with low number of bugs in the dataset. Precisely, we consider a total of 40 bugs, reported for the following 3 projects: Cli (22), Collections (2) and Csv (16). 

For the assertion assessment analysis, we use the dataset from SpecFuzzer, a specification inference technique recently introduced by Molina et al.~\cite{MolinaDA2022}, that includes (41) assertions manually written by developers. 
Each subject contains the source code, the test suite used during the inference process, and the set of manually written expected assertions. 
Particularly, we focus on 4 methods of the dataset (StackAr.pop, StackAr.topAndPop, Angle.getTurn and Composite.addChild) in which 6 assertions from the ground-truth are discarded since they do not kill any mutant (cf.~\cite[Table 4]{MolinaDA2022}). 
We study whether {\OurTool} can help SpecFuzzer in selecting the discarded assertions, and compare with PiTest.

\subsection{Experimental Procedure}
To answer RQ1, we start by generating mutants with {\OurTool} and PiTest for the fixed version of each fault. Table~\ref{tab:mutants_generated} summarises the number of mutants generated by the tools. 
Then, we make an objective comparison between the techniques in terms of the number of generated mutants and faults detected. We select minimal test cases, from the developer test suites, that kill the same number of mutants for both tools and check whether they detect the associated real faults or not. This is important since {\OurTool} generates by far less mutants than PiTest. We then, perform a cost-effective analysis by simulating a scenario where a tester selects mutants based on which he designs tests to kill them. We start by taking the set of mutants created by a tool, randomly picking up a mutant and selecting a test that kills it or judging the mutant as equivalent and discard it. We then run this test with all mutants in the set and discarding those that are killed. We repeat this process until we reach a maximum number of mutants killed. We adopt as effort/cost metric the number of times a developer analyses mutants (either these result to a test or not). This means that effort is the number of tests selected plus the number of mutants judged as equivalent. We then check if the generated test suite detect or not the real faults. We repeat this process 100 times to reduce the impact of the random selection of mutants and killing tests on our results. This cost-effective evaluation aims at emphasising the effects of the different mutant generation approaches.

\begin{table}[tp!]
\vspace{-1.0em}
\centering
\caption{Number of (compilable) mutants generated by {\OurTool} and PiTest for each project.}
\begin{tabular}{lrr}
\toprule
\textbf{Project}  &\textbf{\OurTool} & \textbf{PiTest}\\
\midrule
Cli (22 bugs) & 4.282  & 19.482\\
Collections (2 bugs) & 280  & 1.162\\
Csv (16 bugs)  & 4.515 & 18.378\\
\midrule
\textbf{Total} & \textbf{9.077} & \textbf{39.022}\\
\bottomrule
\end{tabular}
\label{tab:mutants_generated}
\vspace{-1.6em}
\end{table}

To answer RQ2, we start by generating mutants with {\OurTool} and PiTest for the four methods under analysis. Then we run the inference tool, SpecFuzzer~\cite{MolinaDA2022}, to obtain the a set of valid assertions for the method of interest (i.e., never falsified by the test suite). 
SpecFuzzer then performs a mutation analysis on the inferred assertions, and discards the ones that do not kill any mutant. We confirm that the 6 assertions from the ground-truth are discarded in this process. 
Hence, we run again the mutation analysis of SpecFuzzer, but in this case we consider mutants from {\OurTool} and PiTest, and analyse whether the ground-truth assertions are discarded or not.

To answer RQ3, we discuss on some examples from {\OurTool} and the potential benefits that it can provide to mutation testing and assertion inference approaches.

\subsection{Implementation}
{\OurTool} uses Spoon\footnote{\url{https://spoon.gforge.inria.fr}} for manipulating the Java programs. It employs the current pre-trained version of CodeBERT\footnote{\url{https://github.com/microsoft/CodeBERT}}, and provides the scripts to integrate other pre-trained language models if required. 
The source code, a set of examples, and the results of our preliminary evaluation are publicly available at: \textbf{\url{https://github.com/rdegiovanni/mBERT}}.

\begin{figure*}[t]
\vspace{-0.8em}
\centering
\begin{subfigure}[b]{0.32\textwidth}
\centering
\includegraphics[width=.8\textwidth]{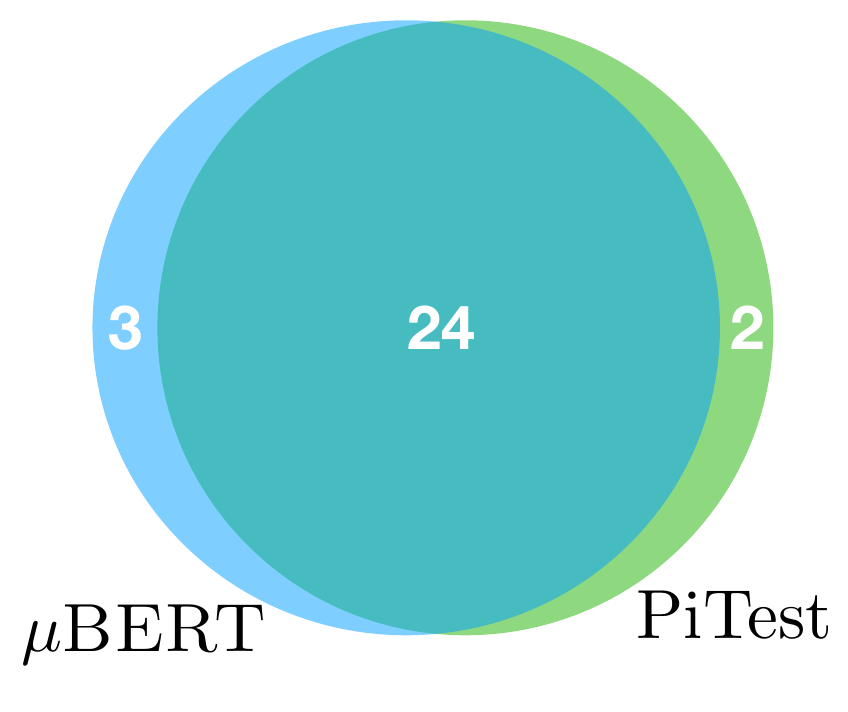}
\caption{{\OurTool} detects 27 out of 40 faults (67.5\%), while PiTest detects 26 (65.0\%). {\OurTool} detects 3 faults not detected by PiTest, but misses 2 faults detected by PiTest. A total of 11 faults out of 40 (27.5\%) were not detected neither by {\OurTool} and PiTest.}
\label{fig:bugs-detected-per-tool}
\end{subfigure}
\hspace{0.8mm}
\begin{subfigure}[b]{0.32\textwidth}
\centering
\includegraphics[width=\textwidth]{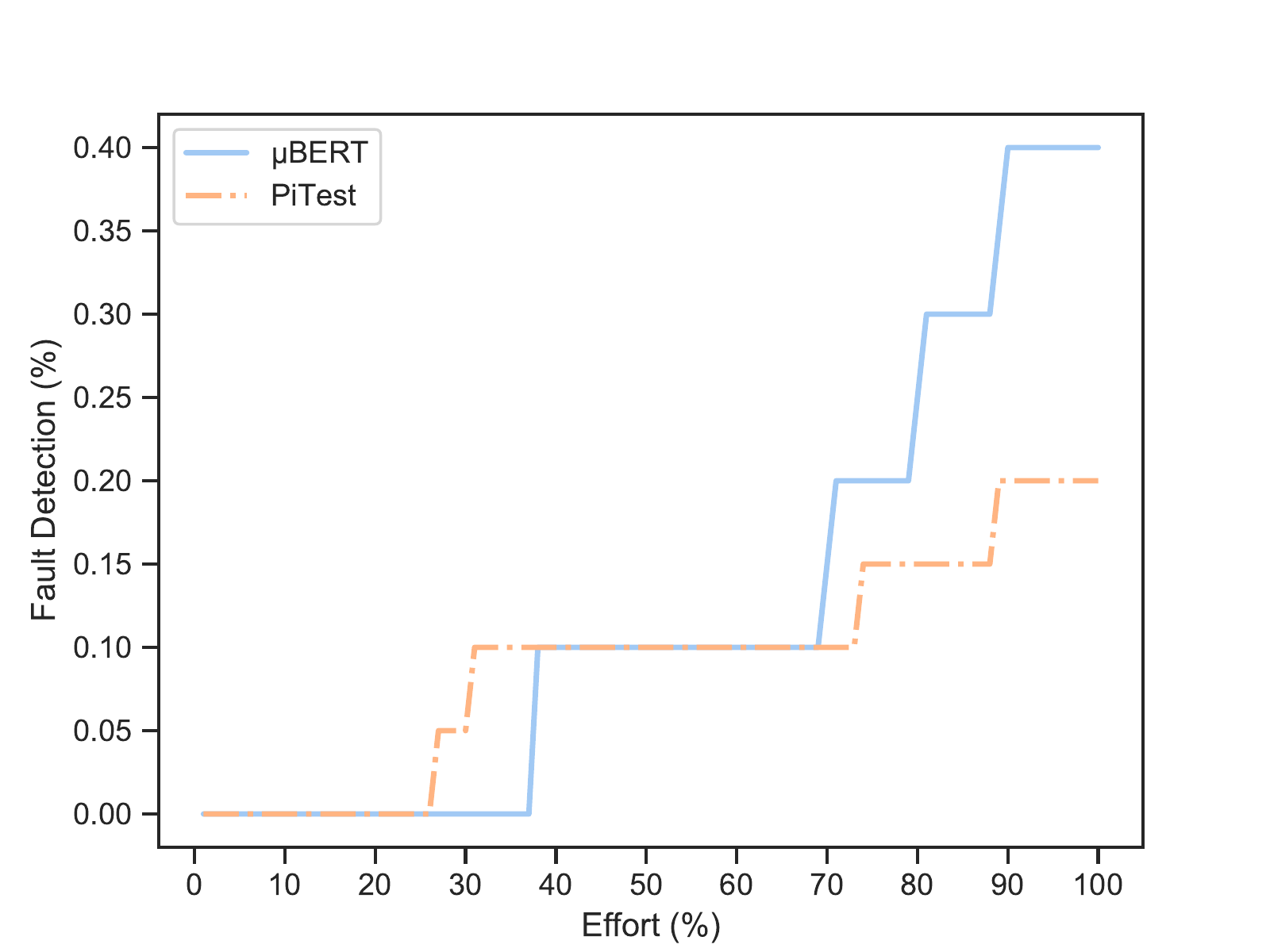}
\caption{Effort (x-axis) indicates the number of mutants analysed by a tester, while the effectiveness (y-axis) indicates the fault detection ratio of the tools. Effort of 100\% means the maximum number of mutants analysed (number of tests and number mutants considered as equivalent) by a tester when using {\OurTool}. 
}
\label{fig:cost-benefit}
\end{subfigure}
\hspace{0.8mm}
\begin{subfigure}[b]{0.32\textwidth}
\centering
\includegraphics[width=\textwidth]{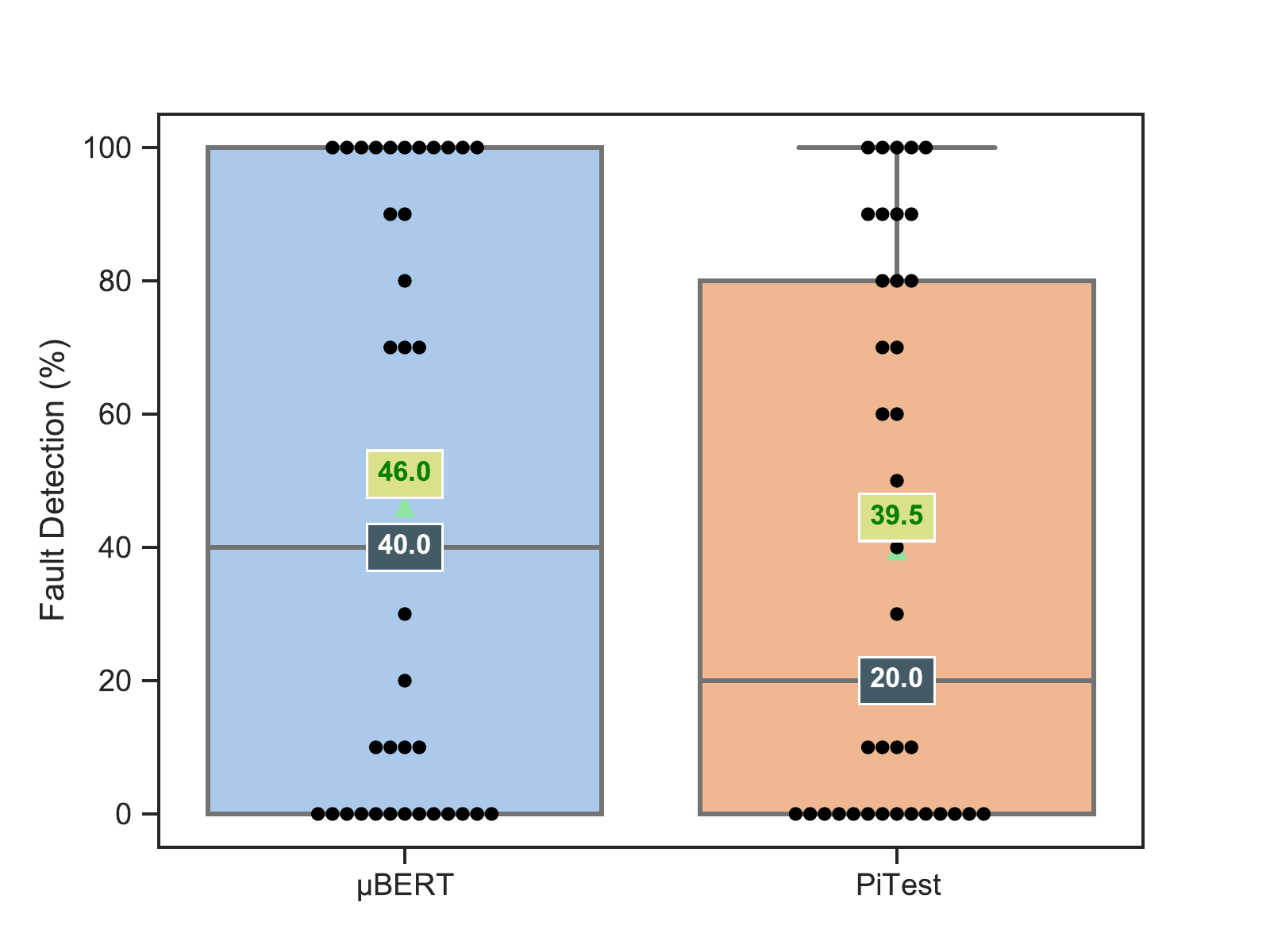}
\vspace{-1.4em}
\caption{Test suites killing all the mutants from {\OurTool} have 40.0\% (median) of likelihood of detecting a real fault (46.0\% in average); while suites killing exactly the same number of PiTest mutants have 20.0\% (median) of likelihood in succeeding (39.5\% in average).
}
\label{fig:revailing_bugs-same-effort}
\end{subfigure}
\caption{\small RQ1: Fault detection comparison between the {\OurTool} and PiTest.}
\label{fig:RQ1-tools-comparison}
\vspace{-0.8em}
\end{figure*}

\section{Experimental Results}
\subsection{RQ1: Fault Detection Analysis}
Figure~\ref{fig:RQ1-tools-comparison} summarises the fault detection capabilities of {\OurTool} and PiTest. 
Figure~\ref{fig:bugs-detected-per-tool} shows that test suites killing \textit{all the mutants from {\OurTool}} can detect 27 out of 40 faults (67.5\%). 
While suites killing \textit{all PiTest mutants} can detect 26 out of 40 faults (65.0\%). 
There are 11 faults (27.5\%) not detected neither by {\OurTool} and PiTest.
When we check for overlapping, we observe that 3 faults detected by {\OurTool} were not detected by PiTest, and 2 faults detected by PiTest were not detected by {\OurTool}. 
These indicate {\OurTool}'s fault detection effectiveness is comparable with the one of PiTest, and {\OurTool} mutants can potentially complement other mutation testing techniques. 

Figure~\ref{fig:cost-benefit} summarises the cost-effective evaluation of the techniques; fault detection effectiveness (y axis) in relation to the same number of analysed mutants (effort) (x axis). An effort of 100\% means that the maximum possible number of mutants were analysed (for {\OurTool}), which in the case of PiTest is the same number as by {\OurTool} to enable a fair comparison. As Table~\ref{tab:mutants_generated} noted, PiTest produces way many more mutants than {\OurTool} and thus killing all its mutants requires way more effort than {\OurTool}. We observe that {\OurTool} is more cost-effective, indicating that suites selected based on the mutants of {\OurTool} are more likely to find real faults than those selected by PiTest, when the same number of mutants are analysed. 
Figure~\ref{fig:revailing_bugs-same-effort} emphasises this cost-effective comparison, and particularly focus the fault detection ratio when the maximum number of mutants was analysed (i.e., total number of mutants generated by {\OurTool}). 
In average (mean), test suites killing all the mutants from {\OurTool} have 40.0\% (46.0\%) of likelihood of detecting a real fault; while, suites killing exactly the same number of PiTest mutants have 20.0\% (39.5\%).  

\subsection{RQ2: Assertion Assessment Analysis}
Table~\ref{tab:assertion_analysis} summarises the performance of SpecFuzzer when uses the mutants from {\OurTool} and PiTest for selecting the assertions. For each tool, we report if the assertions in the ground-truth were selected or discarded (Suc. column), we also report the number of  generated and killed mutants (\#M and \#K, respectively). 
We can observe that 3 out of the 6 assertions under analysis, kill some mutant produced by {\OurTool} and thus, SpecFuzzer does not discard them. 
In the case of PiTest, it helps in preserving 2 out of 6 assertions from the ground-truth, but in general in produces many more mutants than {\OurTool} (e.g., up to 10 times in StackAr program) what affects the time require for filtering the assertions. 

\begin{table}[tp!]
\vspace{-1.0em}
\centering
\caption{Manually written assertions discarded by SpecFuzzer, because they do not kill any mutant~\cite[Tablle 4]{MolinaDA2022}. 
When SpecFuzzer uses the mutants generated by {\OurTool}, it does not discard 3 out of the 6 valid assertions. When it uses PiTest mutants, it preserves 2 out of the 6 assertions, but it analyses many more mutants (up to 10 times).}
\resizebox{\columnwidth}{!}{
\begin{tabular}{l|l|rrr|rrr}
\toprule
 &  &\multicolumn{3}{c|}{\textbf{{\OurTool}}}&\multicolumn{3}{c}{ \textbf{PiTest}}\\
\textbf{Subject} & \textbf{Assertions} &Suc.&\#M&\#K&Suc.&\#M&\#K\\
\midrule
\texttt{StackAr.pop}        & \texttt{theArray[old(top)] == null} &  & 4 & 4 & & 42 & 29 \\
\texttt{StackAr.topAndPop}  & \texttt{theArray[old(top)] == null} &  & 6 & 6 & & 46 & 39 \\
\texttt{Angle.getTurn}      & \texttt{abs(res) <= 1}            & $\checkmark$ & 23 & 23 & $\checkmark$  & 81 & 15 \\
\texttt{Composite.addChild} & \texttt{c.value == old(c.value)} & $\checkmark$ & 86 & 42 & $\checkmark$ & 96 & 52 \\
                            & \texttt{children == old(children)} & $\checkmark$ & & & & &\\
                            & \texttt{ancestors == old(ancestors)} & & & & & &\\
\bottomrule
\end{tabular}
}
\label{tab:assertion_analysis}
\vspace{-1.6em}
\end{table}

\subsection{RQ3: Qualitative Analysis of {\OurTool} Mutants}
Table~\ref{tab:real-faults-mutants} shows examples of mutants produced by {\OurTool} that help in finding the three real faults (namely, faults with ids Cli\_10, Csv\_15 and Csv\_16) not found by PiTest. 
For each case, we report the diff between the fixed and the buggy version, as well as, the diff between the fixed version and the mutants generated by {\OurTool}. 
Lines in {\color{red!60!black}red} correspond to the fixed version, while lines in {\color{green!60!black}green} correspond to the buggy version and the mutants.

The real fault denoted by Cli\_10, located in file Parser.java, resides inside function \texttt{setOptions} and the problem is that it creates an aliasing between the internal object field \texttt{requiredOptions} and same field from the object \texttt{options} given as parameter. 
{\OurTool} generates mutants that interact with this field trough the getter method \texttt{getRequiredOptions()}. 
For instance, MUTANT 1 changes an if condition regarding the size of list containing the required options. MUTANT 2 changes method call \texttt{remove} by \texttt{add}, then the list \texttt{requiredOptions} will add an element instead of removing it.

Csv\_15 is a real fault inside method \texttt{printAndQuote}, located in class CSVFormat.java, in which some chars in the sequence to print were causing a failure in the parser. 
{\OurTool} generates mutants that change predefined special tokens, later used to print the strings. For instance, MUTANT 3 changes the return value of function \texttt{getDelimiter()} by returning always 0, instead of the preset delimiter token.  
MUTANT 4 replaces object \texttt{value} with object \texttt{this} when calling \texttt{toString} in a condition that initialises the values to print.

Fault denoted as Csv\_16 is present in file CVSParser.java, precisely inside class \texttt{CSVRecordIterator} that implements an iterator that returns the records of the csv. 
{\OurTool} generates mutants that change the control flow of the program, for instance, the mutated expression \texttt{this.current == current} in MUTANT 5, will always evaluate to true, and MUTANT 6 introduces an infinite recursion in function \texttt{isClosed}.
MUTANT 7 modifies the initialization of variable \texttt{inputClean} in method \texttt{addRecordValue}, that is later used by the iterator.

If the reader prefer, please refer to the appendix to find more examples of mutants generated by {\OurTool} useful for detecting these faults.

\begin{table*}[tp!]
\vspace{-1.0em}
\centering
\caption{Examples of ``good'' mutants generated by {\OurTool} that help in detecting the faults for Cli\_10, Csv\_15 and Csv\_16, not found by PiTest.}
\begin{tabular}{l}
\toprule
{\textbf{BugID:} Cli\_10. \hspace*{1cm} \textbf{Class:} Parser.java} \\
\midrule
\begin{lstlisting}[language=diff]
@@ PATCH  -44,7 +43,7 @@
- this.requiredOptions = new ArrayList(options.getRequiredOptions());
+ this.requiredOptions = options.getRequiredOptions();

@@ MUTANT 1: -306,7 +306,7 @@ 
- if (getRequiredOptions().size() > 0)
+ if (getRequiredOptions().size() > 1)

@@ MUTANT 2: -402,7 +402,7 @@ 
- getRequiredOptions().remove(opt.getKey());
+   getRequiredOptions().add(opt.getKey());
\end{lstlisting}
\\
\midrule
{\textbf{BugID:} Csv\_15. \hspace*{1cm} \textbf{Class:} CSVFormat.java} \\
\midrule
\begin{lstlisting}[language=diff]
@@ PATCH -1186,7 +1186,9 @@ 
- if (c <= COMMENT) {
+ if (newRecord && (c < 0x20 || c > 0x21 && c < 0x23 || c > 0x2B && c < 0x2D || c > 0x7E)) {
+    quote = true;
+ } else if (c <= COMMENT) {

@@ MUTANT 3: -763,7 +763,7 @@ 
     public char getDelimiter() {
-        return delimiter;
+        return 0;
     }
     
@@ MUTANT 4: -1081,7 +1081,7 @@
- charSequence = value instanceof CharSequence ? (CharSequence) value : value.toString();
+ charSequence = value instanceof CharSequence ? (CharSequence) value : this.toString();
\end{lstlisting}
\\
\midrule
{\textbf{BugID:} Csv\_16. \hspace*{1cm} \textbf{Class:} CSVParser.java} \\
\midrule
\begin{lstlisting}[language=diff]
@@ PATCH
@@ -286,7 +286,6  -355,7 +354,6  -522,10 +520,7 -573,6 +568,7 @@
-    private final CSVRecordIterator csvRecordIterator;
-        this.csvRecordIterator = new CSVRecordIterator();
     public Iterator<CSVRecord> iterator() {
-        return csvRecordIterator;
-    }
-    
-    class CSVRecordIterator implements Iterator<CSVRecord> {
+        return new Iterator<CSVRecord>() {
             private CSVRecord current;
             private CSVRecord getNextRecord() {
                 throw new UnsupportedOperationException();
             }
         };
+    }

@@ MUTANT 5: -542,7 +542,7 @@
-   if (this.current == null) {
+   if (this.current == current) {

@@ MUTANT 6: -505,7 +505,7 @@ 
     public boolean isClosed() {
-        return this.lexer.isClosed();
+        return this.isClosed();
     }             

@@ MUTANT 7: -363,7 +363,7 @@ 
-   final String inputClean = this.format.getTrim() ? input.trim() : input;
+   final String inputClean = "";
\end{lstlisting}
\\
\bottomrule
\end{tabular}
\label{tab:real-faults-mutants}
\vspace{-1.6em}
\end{table*}

Table~\ref{tab:specfuzzer-mutants} shows the mutants generated by {\OurTool} that help to SpecFuzzer to not discard good assertions, taken from the ground-truth. 
Particularly, the 3 mutants created for method $\texttt{Angle.getTurn}$ clearly violate the assertion \texttt{abs(res) <= 1} and thus, it will not be discarded. 

In the case of \texttt{Composite.addChild} we can observe that MUTANT 4 replaces the invocation \texttt{c.setParent(this)} by \texttt{c.update(this)}. This mutant makes that the value of the child object c (\texttt{c.value}) be updated with the value of object \texttt{this} (the parent). Then, assertion \texttt{c.value == old(c.value)} will be clearly violated by this mutant and thus, will not be discarded by SpecFuzzer. 

Similarly, MUTANT 5 replaces invocation \texttt{ancestors.add(p)} by \texttt{children.add(p)}. This mutant clearly can change \texttt{children} set values. Assertion \texttt{children == old(children)} clearly kills this mutant, so SpecFuzzer will preserve it.

\begin{table}[tp!]
\vspace{-1.0em}
\centering
\caption{{\OurTool} generates these mutants that are killed by the ground-truth assertions and thus, SpecFuzzer does not discard them.}
\begin{tabular}{l}
\toprule
\textbf{Subject:} $\texttt{Angle.getTurn}$ \\
\textbf{Assertion:}  \texttt{abs(res) <= 1} \\
\midrule
\begin{lstlisting}[language=diff]
@@ MUTANT 1: -43,7 +43,7 @@ 
    if (crossproduct > 0) {
-       res = 1;
+       res = 2;
              
@@ MUTANT 2: -43,7 +43,7 @@ 
    if (crossproduct > 0) {
-       res = 1;
+       res = 255;
              
@@ MUTANT 3: -43,7 +43,7 @@ 
    if (crossproduct > 0) {
-       res = 1;
+       res = 360;
\end{lstlisting}
\\
\midrule
\textbf{Subject:} $\texttt{Composite.addChild}$ \\
\textbf{Assertion:} \texttt{c.value == old(c.value)} \\
\midrule
\begin{lstlisting}[language=diff]
@@ MUTANT 4: @@ -70,7 +70,7 @@ 
-        c.setParent(this);
+        c.update(this);
\end{lstlisting}
\\
\midrule
\textbf{Subject:} $\texttt{Composite.addChild}$\\
\textbf{Assertion:} \texttt{children == old(children)} \\
\midrule
\begin{lstlisting}[language=diff]
@@ MUTANT 4: @@ -82,7 +82,7 @@ 
-        ancestors.add(p);
+        children.add(p);
\end{lstlisting}
\\
\bottomrule
\end{tabular}
\label{tab:specfuzzer-mutants}
\vspace{-1.6em}
\end{table}

\section{Threats to Validity}
\label{sec:threats-to-validity}
One of the threats related to \emph{external validity} relies on the election of the projects from Defects4J used in our  evaluation (Cli, Collections and Csv).  
This is a preliminary study and we do not exclude the threat of having different results when conducting the same study on other projects from other domains.
Other threat is related to the use of the mutation testing tool PiTest as a baseline in our experiments. Despite that this is one of the state-of-the-art tools for creating mutants, the results may change when compared with other mutant generation techniques. 

\emph{Internal validity} threats may relate with our implementation of {\OurTool}. To mitigate this threat we made publicly available our implementation, repeated several times the experiments, and manually validated the results. 
Other threat may arise from the type of expressions selected to mutate (mutation operators), whose effectiveness can be affected when applied to other projects, or implemented in other programming language. 
To mitigate this threat, {\OurTool} mutates expressions typically handled by mutation testing tools, such as PiTest, and it is also possible to extend our implementation to provide further mutation operators if required.  
The performance of CodeBERT can also affect {\OurTool}'s effectiveness. Currently, {\OurTool} uses CodeBERT as a black-box, so it can be benefit for future improvements of the pre-trained model. Moreover, generated mutants may change if a different pre-trained model is employed for predicting the masked tokens.

Regarding construct validity threats, our assessment metrics, such as the number of mutants analysed and number of faults found, may not reflect the actual testing cost~/~effectiveness values. 
However, these metrics have been widely used by the literature \cite{PapadakisK00TH19,AndrewsBLN06,KurtzAODKG16} and are intuitive, since the number of analyzed mutants essentially simulate the manual effort involved by testers, while the test suites selected to kill the mutants can also be used to measure its effectiveness in finding the fault. 
In our experiments, test cases were selected from the pool of tests provided by Defects4J, which may not reflect the real cost/effort in designing such test cases.

\section{Related Work}

Mutation testing has a long history with multiple advances \cite{PapadakisK00TH19}, either on the faults that it injects or on the processes that it supports. Despite the rich history, the creation of ''good'' mutants is a question that remains. 

The problem has traditionally been addressed by the definition of mutation operators using the underlying programming language syntax. These definitions span across languages \cite{DelamaroMM01, MaKO02, DengOAM17}, artefacts (such as specification languages and behavioural models) 
\cite{KrennSTAJB15, PapadakisHT14, HieronsM09}, and specialised applications (such as energy-aware \cite{JabbarvandM17} and security-aware \cite{LoiseDPPH17} operators). 

More recent attempts include the composition of mutation operators (composition of fault patterns) using historical fault fixing commits. These approaches are either mined using simple syntactic changes \cite{BrownVLR17}, or more complex patterns manually crafted \cite{khanfir2020ibir}, or automatically crafted patterns using machine translation techniques \cite{tufano2019learning}. 

Independently of the way mutants are created, they are often too many to be used, with many of them being of different "quality" \cite{PapadakisCT18}, as they are either trivially killed or simply redundant. To this end, recent attempts aim at selecting mutants that are likely killable \cite{ChekamPBTS20, PeacockDDC21, Duque-TorresDPR20}, likely to couple with real faults \cite{ChekamPBTS20}, likely subsuming \cite{GODCPT, JuniorDDSVD20, GheyiRSGFdATF21} or relevant to regression changes  \cite{ma2021mudelta}.  

Our notion of mutant naturalness is somehow similar to the n-gram based notion of naturalness used by  Jimenez et al. \cite{JimenezCCPKTH18}. Though,  we differ as we generate mutants instead of selecting and rely on a transformer-based neural architecture that captures context both before and after the mutated point.

\section{Conclusion and Future Work}
We presented {\OurTool}, a mutation testing approach that generates ``natural'' mutants by leveraging self-supervised model pre-training of big code. As such it does not require any training on historical faults, or other mutation testing data that are expensive to gather, but rather large corpus of source code that are easy to gather and use. Interestingly, our analysis showed that {\OurTool}'s performance is comparable with traditional mutation testing  tools, and even better in some cases,  both in terms of fault detection and assertion inference. These results suggests that ``natural'' mutants do not only concern readability but also test effectiveness. Perhaps more importantly, {\OurTool} is the first attempt that leverage self-supervised language methods in mutation testing, thereby opening new directions for future research. 

There are a few lines of future work that we plan to explore. 
We plan to extend our evaluation to the entire datasets of  Defects4J and SpecFuzzer for analysing {\OurTool}'s fault detection and assertion inference capabilities. 
We also plan to include other mutation testing tools than PiTest in the comparison. 
So far {\OurTool} uses CodeBERT as a black-box and mutants are generated in a brute-force way, i.e., we iterate on every program statement to mask and generate mutants. 
We plan to analyse CodeBERT's embedding and predictions to study whether it is possible to predict ``interesting'' locations to mutate, for instance, locations where \emph{subsuming mutants} can be generated~\cite{ammann_establishing_2014}.

\section*{Acknowledgment}

This work is supported by the Luxembourg National Research Funds (FNR) through the INTER project grant, INTER/ANR/18/12632675/SATOCROSS.

\bibliographystyle{plain}
\bibliography{bibfile}

\onecolumn
\appendix[Extra examples]

\begin{table*}[!htp]
\centering
\caption{This table presents more mutants generated by {\OurTool} for the faults Cli\_10, Csv\_15 and Csv\_16.}
\begin{tabular}{l}
\toprule
{\textbf{BugID:} Cli\_10. \hspace*{1cm} \textbf{Class:} Parser.java} \\
\midrule
\begin{lstlisting}[language=diff]
@@ MUTANT: -306,7 +306,7 @@ 
- if (getRequiredOptions().size() > 0)
+ if (getRequiredOptions().size() > 2)

@@ MUTANT: -321,7 +321,7 @@
- throw new MissingOptionException(buff.substring(0, buff.length() - 2));
+ throw new MissingOptionException(buff.substring(0, buff.length()+2));
\end{lstlisting}
\\
\midrule
{\textbf{BugID:} Csv\_15. \hspace*{1cm} \textbf{Class:} CSVFormat.java} \\
\midrule
\begin{lstlisting}[language=diff]
@@ MUTANT: -790,7 +790,7 @@ 
public String[] getHeaderComments() {
-        return headerComments != null ? headerComments.clone() : null;
+        return headerComments==null ? headerComments.clone() : null;
}

@@  MUTANT: -879,7 +879,7 @@ 
    public boolean getTrailingDelimiter() {
-       return trailingDelimiter;
+       return true;
  }

@@ MUTANT: -1081,7 +1081,7 @@ 
- charSequence = value instanceof CharSequence ? (CharSequence) value : value.toString();
+ charSequence = value instanceof Object ? (CharSequence) value : value.toString();

@@ MUTANT: -1726,7 +1726,7 @@ 
 return new CSVFormat(delimiter, quoteCharacter, quoteMode, commentMarker, escapeCharacter,
- ignoreSurroundingSpaces, ignoreEmptyLines, recordSeparator, nullString, headerComments, header,
+ ignoreSurroundingSpaces, ignoreEmptyLines, null, nullString, headerComments, header,
   skipHeaderRecord, allowMissingColumnNames, ignoreHeaderCase, trim, trailingDelimiter, autoFlush);
\end{lstlisting}
\\
\midrule
{\textbf{BugID:} Csv\_16. \hspace*{1cm} \textbf{Class:} CSVParser.java} \\
\midrule
\begin{lstlisting}[language=diff]
@@ MUTANT: -362,7 +362,7 @@ 
-        final String input = this.reusableToken.content.toString();
+        final String input = this.toString();

@@ MUTANT: -557,7 +557,7 @@ 
-            if (next == null) {
+            if (current == null) {

@@ MUTANT: -363,7 +363,7 @@ 
-        final String inputClean = this.format.getTrim() ? input.trim() : input;
+        final String inputClean = this.format.getTrim() ? input.trim() : "";

@@ MUTANT: -463,7 +463,7 @@ 
-        if (formatHeader != null) {
+        if (format != null) {

@@ MUTANT: -463,7 +463,7 @@ 
-        if (formatHeader != null) {
+        if (this != null) {

@@ MUTANT: -546,7 +546,7 @@ 
-            return this.current != null;
+            return this != null;
\end{lstlisting}
\\
\bottomrule
\end{tabular}
\label{tab:extra-mutants}
\end{table*}

\end{document}